\documentclass[aps,prd,nofootinbib,preprintnumbers,floatfix,notitlepage,longbibliography]{revtex4-2}

\usepackage{amsmath,amsthm,amssymb}
\usepackage{hyperref}

\begin{document}

\title{Gravitational-wave polarizations in generic linear massive gravity and generic higher-curvature gravity}

\author{Tomoya Tachinami}\email[]{tachinami(a)tap.st.hirosaki-u.ac.jp}
\affiliation{
Graduate School of Science and Technology, Hirosaki University,
Hirosaki, Aomori 036-8561, Japan
}
\author{Shinpei Tonosaki}\email[]{tonosaki(a)tap.st.hirosaki-u.ac.jp}
\affiliation{
Graduate School of Science and Technology, Hirosaki University,
Hirosaki, Aomori 036-8561, Japan
}
\author{Yuuiti Sendouda}\email[]{sendouda(a)hirosaki-u.ac.jp}
\affiliation{
Graduate School of Science and Technology, Hirosaki University,
Hirosaki, Aomori 036-8561, Japan
}

\date{\today}

\begin{abstract}
We study the polarizations of gravitational waves (GWs) in two classes of extended gravity theories.
As a preparatory yet complete study, we formulate the polarizations in linear massive gravity (MG) with generic mass terms of non-Fierz--Pauli type by identifying all the independent variables that obey Klein--Gordon-type equations.
The dynamical degrees of freedom (dofs) in the generic MG consist of spin-$ 2 $ and spin-$ 0 $ modes, the former breaking down into two tensor (helicity-$ 2 $), two vector (helicity-$ 1 $) and one scalar (helicity-$ 0 $) components, while the latter just corresponding to a scalar.
We find convenient ways of decomposing the two scalar modes of each spin into distinct linear combinations of the transverse and longitudinal polarizations with coefficients directly expressed by the mass parameters, thereby serving as a useful tool in measuring the masses of GWs.
Then we analyze the linear perturbations of generic higher-curvature gravity (HCG) whose Lagrangian is an arbitrary polynomial of the Riemann tensor.
When expanded around a flat background, the linear dynamical dofs in this theory are identified as massless spin-$ 2 $, massive spin-$ 2 $, and massive spin-$ 0 $ modes.
As its massive part encompasses the identical structure to the generic MG, GWs in the generic HCG provide six massive polarizations on top of the ordinary two massless modes.
In parallel to MG, we find convenient representations for the scalar-polarization modes directly connected to the parameters of HCG.
In the analysis of HCG, we employ two distinct methods;
One takes full advantage of the partial equivalence between the generic HCG and MG at the linear level, whereas the other relies upon a gauge-invariant formalism.
We confirm that the two results agree.
We also discuss methods to determine the theory parameters by GW-polarization measurements.
It is worth stressing that our method for determining the theory parameters does not require measuring the propagation speeds, whether absolute or relative, or the details of the waveforms of the GWs.
\end{abstract}

\maketitle

\section{Introduction}

In the era of gravitational-wave astronomy opened by the historic event GW150914 \cite{Abbott:2016blz}, increasingly more attention has been attracted to the nature of the dynamical degrees of freedom (dofs) of gravity propagating as gravitational waves (GWs).
Detection of any deviations from the predictions of general relativity (GR), exactly luminal propagation and two orthogonal modes, would immediately signal the presence of the gravitational theory beyond GR.

A representative example of extended gravity theories is the linear massive gravity (MG), in which gravitational waves acquire masses.
Adding a generic combination of possible two mass terms to GR breaks its four linear gauge symmetries and gives rise to six dofs in total.
A special class of MG introduced by Fierz and Pauli (FP) \cite{Fierz:1939ix} with a single mass parameter $ m $ is known to avoid the appearance of the spin-$ 0 $ mode that would have a negative kinetic term, see e.g., \cite{Hinterbichler:2011tt} for a review.

If GWs are massive, their velocity $ c_\mathrm g $ deviates from the speed of light $ c $ due to the modification of the dispersion relation.
The multi-messenger analysis of the GW event GW170817 \cite{TheLIGOScientific:2017qsa} has put a tight constraint on the deviation, $ |c_\mathrm g-c|/c \lesssim \mathcal O(10^{-15}) $ \cite{Monitor:2017mdv}, which can be interpreted as an upper limit on the mass.
It was also pointed out that a non-zero mass, however small, of graviton in the FP theory would lead to a bending angle of light around a massive body discontinuously different from GR \cite{Iwasaki:1971uz,vanDam:1970vg,Zakharov:1970cc}, which could, however, be circumvented by nonlinearities \cite{Vainshtein:1972sx}.

It was argued in \cite{Eardley:1973br,Eardley:1974nw} that GWs in a certain generic class of gravity theories can have maximally six polarizations.
Although there has been no contradiction with the hypothesis of only two polarizations in the GW experiments \cite{Abbott:2017tlp} including the observed orbital decay rate of a neutron star binary PSR B1913+16 \cite{1982ApJ...253..908T}, the ongoing progress in the construction of the worldwide GW-observatory network has motivated the developments of various methods for detecting those beyond-Einstein polarization modes \cite{Hayama:2012au,Isi:2015cva,Isi:2017equ,Takeda:2018uai}.

In practice, a complete decomposition into possible six polarizations and determination of each amplitude in an observed GW signal cannot be done with the limited number of detectors that we currently have, so there have been only weak constraints on the existence of non-GR polarizations;
An example is the constraint on the vector-type polarizations obtained by a method to detect scalar and vector polarizations with four interferometers of LIGO (Hanford and Livingston), Virgo and KAGRA developed in \cite{Hagihara:2018azu,Hagihara:2019ihn,Hagihara:2019rny}.
In this regards, there are still open and wide theoretical possibilities to explore.

One of our purposes in this paper is to study gravitational-wave polarizations in gravity theories whose Lagrangian can be written in a generic form $ L = f(R^\mu{}_{\nu\rho\sigma},g_{\mu\nu})/2\kappa $\,, where $ g_{\mu\nu} $ is the space-time metric, $ R^\mu{}_{\nu\rho\sigma} $ the Riemann tensor and $ \kappa $ the bare gravitational constant.
The scalar function $ f $ is almost generic but we here conservatively assume that, when Taylor expanded around $ R^\mu{}_{\nu\rho\sigma} = 0 $, it only gives positive powers of the Riemann tensor so that Minkowski space-time is a solution of the full theory.
Studies of such models date back to Weyl \cite{Weyl:1919fi}, who suggested $ f = C_{\mu\nu\rho\sigma}\,C^{\mu\nu\rho\sigma} $\,, where $ C^\mu{}_{\nu\rho\sigma} $ is the Weyl curvature tensor.
Relatively modern motivations also come from the developments in string theories, e.g., \cite{Gross:1986iv}.
Einstein's general relativity is defined by the linear function $ f = R $\,, while presence of any higher-order terms characterizes how the theory differs from GR.

Higher-curvature gravity (HCG) generically exhibits more dynamical dofs than GR does.
For instance, the theory with $ f = R + \beta\,R^2 $ can be shown to be equivalent to a scalar--tensor theory, which can be generalized to the case of a generic function $ f $ of the Ricci scalar, called $ f(R) $ gravity.
As a cosmological application, the $ R + \beta\,R^2 $ model was utilized by Starobinsky to realize inflation \cite{Starobinsky1980}.
A generic class of $ f(R) $ gravity has provided candidates for the dark energy although such theories with negative powers of curvature is out of our scope in this paper.
Also, it was shown by Stelle \cite{Stelle:1977ry} that in the theory with $ f = R - \alpha\,C_{\mu\nu\rho\sigma}\,C^{\mu\nu\rho\sigma} + \beta\,R^2 $\,, there arises another massive spin-$ 2 $ particle on top of zero-mass graviton, which was utilized to render the quantum theory renormalizable \cite{Stelle:1976gc}.
Afterwards, a general Hamiltonian analysis of $ f(\text{Riemann}) $ gravity, keeping $ f $ undetermined, was done in \cite{Deruelle:2009zk}.

Our goal is to give a complete classification of GWs in generic higher-curvature theories in terms of six types of polarizations.
In the literature, some special cases have been studied.
For instance, Bogdanos \textit{et al.}~\cite{Bogdanos:2009tn} considered a generic theory whose Lagrangian consists of quadratic scalar invariants, showing that the GWs have full six polarizations.
More recently, an analysis of $ f(R) $ gravity using gauge-invariant variables was done by Moretti \textit{et al.}~\cite{Moretti:2019yhs}.
However, no complete studies have been conducted on the case of generic $ f $ including the Weyl and Ricci tensors.

In the studies of linear perturbations of HCG, what is significant would be the equivalence between the quadratic curvature gravity and massive bigravity at the linear-order level \cite{Stelle:1977ry}, which was recently extended to arbitrary background with an Einstein metric \cite{Niiyama:2019fvf}.
This means that the analyses of the extra massive dofs in HCG can be done in parallel to MG, which motivates us to start with studying MG in this paper.
On the other hand, we show that a dedicated analysis based on a gauge-invariant formalism is also useful, and that the two results agree.

Specifically, linearly perturbed generic HCG incorporates the structure of linear MG with a generic mass term $ h_{\mu\nu}\,h^{\mu\nu} - (1-\epsilon)\,h^2 $ violating the Fierz--Pauli tuning ($ \epsilon \neq 0 $).
We will analyze the polarizations of GWs in generic MG leaving $ \epsilon $ unspecified so as to generalize the results for specific values of $ \epsilon $ performed, e.g., in \cite{dePaula:2004bc,Corda:2007zz,Corda:2008ki}.

The organization of this paper is as follows.
In Sec.~\ref{sec:pol}, we briefly introduce the basic notion of the polarizations of gravitational waves on the basis of geodesic deviation.
In Sec.~\ref{sec:GW}, we analyze the polarizations of GWs in several theories of gravity:
Starting with reproducing the standard result in general relativity in Sec.~\ref{sec:GR}, we study the generic linear MG and generic HCG in Secs.~\ref{sec:mg} and \ref{sec:hcg}, respectively.
In Sec.~\ref{sec:hcg_bi}, we introduce auxiliary fields with spin-$ 2 $ and spin-$ 0 $ to rewrite the action, count the number of dofs and reveal the origin of six polarizations.
In Sec.~\ref{sec:hcg_gi}, as an another approach, a gauge-invariant formulation to analyze GWs in HCG is developed.
We confirm that the same result is obtained in the two different approaches.
In Sec.~\ref{sec:obs}, we briefly discuss possible methods for determining the theory parameters by GW-polarization observations using laser interferometers or pulsar timing.
Finally, we conclude in Sec.~\ref{sec:concl}.

Throughout the paper, we will work with natural units with $ c = 1 $.
Greek indices of tensors such as $ \mu,\nu,\cdots $ are of space-time while Latin ones such as $ i,j,\cdots $ are spatial.
We introduce background coordinates $ (x^0,x^1,x^2,x^3) = (t,x,y,z) $ in which the Minkowski metric is $ \eta_{\mu\nu} = \mathrm{diag}(-1,1,1,1) $.
The symbol $ \partial_\mu $ denotes partial differentiation $ \frac{\partial}{\partial x^\mu} $\,.
$ \square \equiv \eta^{\mu\nu}\,\partial_\mu \partial_\nu $ is the d'Alembertian and $ \triangle \equiv \delta^{ij}\,\partial_i \partial_j $ the Laplacian.
The Riemann tensor is defined as $ R^\mu{}_{\nu\rho\sigma} = \partial_\rho \Gamma^\mu{}_{\nu\sigma} - \cdots $\,.
Parentheses around tensor indices denote symmetrization such as $ T_{(\mu\nu)} \equiv \frac{1}{2}\,(T_{\mu\nu}+T_{\nu\mu}) $, while square brackets denote antisymmetrization such as $ T_{[\mu\nu]} \equiv \frac{1}{2}\,(T_{\mu\nu}-T_{\nu\mu}) $.

\section{\label{sec:pol}Polarizations of gravitational waves}

In this section, we introduce the notion of polarizations of GWs leaving gravity theory unspecified.
On the microscopic side, the graviton treated as a massless particle with spin $ 2 $ is irreducibly decomposed into the helicity states called \textit{polarizations}.
On the macroscopic side where classical GWs are considered, a way to define \textit{polarization} of GWs is the geodesic deviation in space-times.
To follow the latter perspective, we shall take Minkowskian background $ \eta_{\mu\nu} $ and define the variable as the deviation of the space-time metric $ g_{\mu\nu} $ from the background,
\begin{equation}
h_{\mu\nu}
\equiv
  g_{\mu\nu} - \eta_{\mu\nu}\,.
\end{equation}
Geometric quantities such as curvature are expanded in $ h_{\mu\nu} $\,.
The space-time (space) indices of linear quantities such as the perturbation itself $ h_{\mu\nu} $ are raised and lowered by the background metric $ \eta_{\mu\nu} $ ($ \delta_{ij} $).
Hereafter, the background Laplacian $ \triangle \equiv \delta^{ij} \partial_i \partial_j $ acting on a perturbative quantity is assumed to be invertible.

\subsection{Geodesic deviation and gauge-invariant perturbations}

The principle of detecting gravitational waves with the interferometers like LIGO, Virgo and KAGRA or the pulsar timing arrays is to measure the spatial separation of a nearby, as compared with the wavelength of GWs, pair of test bodies at rest, $ \zeta^i $\,, whose motion is governed by the geodesic deviation equation
\begin{equation}
\ddot\zeta^i
= -{}^{(1)}R^i{}_{0j0}\,\zeta^j\,,
\label{eq:geod}
\end{equation}
where the dot denotes derivative with respect to $ x^0 = t $ and $ {}^{(1)}R^i{}_{0j0} $ is the linear Riemann tensor given in terms of the metric perturbation as
\begin{equation}
{}^{(1)}R_{i0j0}
= -\frac{1}{2}\,\ddot h_{ij}
  + \partial_{(i} \dot h_{j)0}
  - \frac{1}{2}\,\partial_i \partial_j h_{00}\,.
\label{eq:Riemann_h}
\end{equation}
Thus, the separation of two test bodies would fluctuate in response to the gravitational waves embodied by $ h_{\mu\nu} $\,, and conversely, by tracking their movements, one could decode the dynamical contents of the gravitational theory.

Generally, complications in the analysis of gravitational perturbations arise from the intrinsic degrees of freedom of choosing the background coordinates.
The metric perturbations are transformed by a small coordinate change $ x^\mu \to x^\mu + \xi^\mu(x) $ with an arbitrary four vector $ \xi^\mu $ as
\begin{equation}
h_{\mu\nu}
\to
  h_{\mu\nu} - \partial_\mu \xi_\nu - \partial_\nu \xi_\mu\,.
\label{eq:gt}
\end{equation}
An effective way to isolate the physical degrees of freedom is the gauge-invariant formalisms originally developed in the cosmological context \cite{Bardeen:1980kt,Kodama:1985bj,Mukhanov:1990me}.
In this formalism, we begin with introducing scalar, vector and tensor-type variables to decompose each component of the metric perturbation as
\begin{equation}
h_{00}
= -2 A\,,
\quad
h_{0i}
= -\partial_i B - B_i\,,
\quad
h_{ij}
= 2 \delta_{ij}\,C
  + 2 \partial_i \partial_j E
  + 2 \partial_{(i} E_{j)}
  + 2 H_{ij}\,,
\label{eq:h_SVT}
\end{equation}
where the vector and tensor variables satisfy
\begin{equation}
\partial^i B_i
= \partial^i E_i
= 0\,,
\quad
H_i{}^i = 0\,,
\quad
\partial^j H_{ij} = 0\,.
\end{equation}
Gauge transformations of each variable are summarized in Appendix~\ref{sec:gauge}.
We can find a reduced number of variables that are invariant under the transformation \eqref{eq:gt}:
The tensor variable $ H_{ij} $ is invariant;
For the vector part, a combination of the variables $ B_i $ and $ E_i $ which is invariant is
\begin{equation}
\Sigma_i
\equiv
  B_i + \dot E_i\,;
\end{equation}
For the scalar part, a useful set of invariant combinations is
\begin{equation}
\Psi
\equiv
  A - \dot B - \ddot E\,,
\quad
\Phi
\equiv
  C\,.
\end{equation}
At this point, we have found six gauge-invariant variables out of the original ten in $ h_{\mu\nu} $\,.
A remarkable property of the linear Riemann tensor \eqref{eq:Riemann_h} is its invariance under \eqref{eq:gt}.
Indeed, the Riemann tensor can be written in terms of these invariant variables as
\begin{equation}
{}^{(1)}R_{i0j0}
= -\ddot H_{ij}
  - \partial_{(i} \dot\Sigma_{j)}
  + \partial_i \partial_j \Psi
  - \delta_{ij}\,\ddot\Phi\,.
\label{eq:Riemann_gi}
\end{equation}

It is worth stressing that, in spite of their apparent advantage, each gauge-invariant variable does not necessarily correspond to a single dynamical dof in a given gravity theory.
Instead, we generally expect that the gauge-invariant variables become linear combinations of independent dofs.
In particular, a mixture of different spins occurs in the scalar part as we will see in the cases of massive gravity and higher-curvature gravity.

\subsection{Polarization basis}

The connection between the irreducible decomposition of the Riemann tensor \eqref{eq:Riemann_gi} and observables in gravitational-wave experiments is made explicit by considering a wave solution propagating in a fixed direction.
As we already noted, we should keep in mind that each part can contain multiple dofs.

Suppose a tensor variable $ T_{ij}(z-c_T t) $ propagating in the $ z $ direction with velocity $ c_T $\,.
A conventional orthogonal basis that is compatible with the conditions $ T_i{}^i = \partial_i T^{ij} = 0 $ is the $ + $ and $ \times $ polarizations
\begin{equation}
e^+_{ij}
\equiv
\begin{pmatrix}
1 & 0 & 0 \\
0 & -1 & 0 \\
0 & 0 & 0
\end{pmatrix}\,,
\quad
e^\times_{ij}
\equiv
\begin{pmatrix}
0 & 1 & 0 \\
1 & 0 & 0 \\
0 & 0 & 0
\end{pmatrix}\,.
\end{equation}
Using these basis tensors, we can decompose the tensor variable as
\begin{equation}
T_{ij}
= T_{+}\,e^+_{ij} + T_{\times}\,e^\times_{ij}\,,
\end{equation}
where the polarization components are defined as
\begin{equation}
T_\lambda
\equiv
  \frac{1}{2}\,e_\lambda^{ij}\,T_{ij}
\end{equation}
for $ \lambda = +,\times $.
Explicitly, $ T_+ = T_{xx} = -T_{yy} $ and $ T_\times = T_{xy} = T_{yx} $\,.

For a vector variable propagating in the $ z $ direction with velocity $ c_V $\,, $ V_i(z-c_V t) $, a polarization basis compatible with the condition $ \partial_i V^i = 0 $ is
\begin{equation}
e^x_{ij}
\equiv
\begin{pmatrix}
0 & 0 & 1 \\
0 & 0 & 0 \\
1 & 0 & 0
\end{pmatrix}\,,
\quad
e^y_{ij}
\equiv
\begin{pmatrix}
0 & 0 & 0 \\
0 & 0 & 1 \\
0 & 1 & 0
\end{pmatrix}\,.
\end{equation}
Using these, we can decompose the vector-part of the symmetric tensor as
\begin{equation}
\partial_{(i} V_{j)}
= \frac{1}{2}\,V_x'\,e^x_{ij} + \frac{1}{2}\,V_y'\,e^y_{ij}\,,
\end{equation}
where the prime denotes derivative with respect to $ z $.

As for the scalar-type polarizations, natural transverse (``breathing'') and longitudinal polarization bases are
\begin{equation}
e^\mathrm B_{ij}
\equiv
\begin{pmatrix}
1 & 0 & 0 \\
0 & 1 & 0 \\
0 & 0 & 0
\end{pmatrix}\,,
\quad
e^\mathrm L_{ij}
\equiv
\sqrt 2\,
\begin{pmatrix}
0 & 0 & 0 \\
0 & 0 & 0 \\
0 & 0 & 1
\end{pmatrix}\,.
\end{equation}
For a scalar variable propagating in the $ z $ direction with velocity $ c_S $\,, $ S(z-c_S t) $,
\begin{equation}
\left(\partial_i \partial_j - \triangle\,\delta_{ij}\right)\,S
= e^\mathrm B_{ij}\,S''\,,
\quad
\partial_i \partial_j S
= e^\mathrm L_{ij}\,S''\,.
\end{equation}
Sometimes it is also useful to introduce other scalar bases instead of $ \mathrm B $ and $ \mathrm L $, such as
\begin{equation}
e^\mathrm T_{ij}
\equiv
  \sqrt{\frac{2}{3}}\,e^\mathrm B_{ij}
  + \frac{1}{\sqrt 3}\,e^\mathrm L_{ij}
= \sqrt{\frac{2}{3}}\,
\begin{pmatrix}
1 & 0 & 0 \\
0 & 1 & 0 \\
0 & 0 & 1
\end{pmatrix}\,,
\quad
e^\mathbb T_{ij}
\equiv
  \frac{1}{\sqrt 3}\,e^\mathrm B_{ij}
  - \sqrt{\frac{2}{3}}\,e^\mathrm L_{ij}
= \frac{1}{\sqrt 3}
\begin{pmatrix}
1 & 0 & 0 \\
0 & 1 & 0 \\
0 & 0 & -2
\end{pmatrix}\,.
\end{equation}

These basis tensors satisfy an orthonormal condition $ e^\alpha_{ij}\,e^\beta_{ij} = 2 \delta_{\alpha\beta} $ for $ \alpha,\beta = +,\times,x,y,\mathrm B,\mathrm L $.
Seen as a spatial symmetric tensor, the Riemann tensor $ {}^{(1)}R_{0i0j} $ can be decomposed with the above introduced polarization basis.
It is understood from \eqref{eq:geod} that the amplitude of small oscillation of the distance between two bodies $ \delta\zeta^i $ due to the $ \alpha $ polarization component of GW is proportional to $ A_\alpha \equiv \mathcal A\,e_\alpha^{ij}\,{}^{(1)}R_{0i0j} $\,, where $ \mathcal A $ is a constant independent of the type of polarization.

\section{\label{sec:GW}Gravitational-wave polarizations in generic theories}

In this section, we treat concrete examples of gravitational theories: general relativity (GR), generic massive gravity (MG) and generic higher curvature gravity (HCG).

\subsection{\label{sec:GR}General relativity}

We first reproduce the standard result in GR.
The action of GR expanded around a flat background to the quadratic order is 
\begin{equation}
S_\mathrm{GR}[h_{\mu\nu}]
= -\frac{1}{4\kappa}\,\int\!\mathrm d^4x\,{}^{(1)}G_{\mu\nu}\,h^{\mu\nu}\,,
\label{eq:action_gr}
\end{equation}
where $ {}^{(1)}G_{\mu\nu} $ is the linearized Einstein tensor
\begin{equation}
{}^{(1)}G_{\mu\nu}
\equiv
  -\frac{1}{2}\,\square h_{\mu\nu}
  + \partial_{(\mu} \partial^\lambda h_{\nu)\lambda}
  - \frac{1}{2}\,\partial_\mu \partial_\nu h
  + \frac{1}{2}\,\eta_{\mu\nu}\,
    \left(\square h - \partial^\rho \partial^\sigma h_{\rho\sigma}\right)\,.
\label{eq:ein}
\end{equation}
The equation of motion (eom) for $ h_{\mu\nu} $ in vacuum is obtained as
\begin{equation}
{}^{(1)}G_{\mu\nu}
= 0\,.
\label{eq:eom_gr}
\end{equation}
The GR action \eqref{eq:action_gr} and eom \eqref{eq:eom_gr} are invariant under the gauge transformation \eqref{eq:gt}.
We can choose the transverse-traceless (TT) gauge $ h_{\mu\nu} \to h^\mathrm{TT}_{\mu\nu} $ such that
\begin{equation}
h^\mathrm{TT}_{00}
= 0\,,
\quad
h^\mathrm{TT}_{0i}
= 0\,,
\quad
\delta^{ij}\,h^\mathrm{TT}_{ij}
= 0\,,
\quad
\partial^i h^\mathrm{TT}_{ij}
= 0\,.
\end{equation}
Then the eom \eqref{eq:eom_gr} reduces to a massless Klein--Gordon-type equation for $ h^\mathrm{TT}_{ij} $\,:
\begin{equation}
\square h^\mathrm{TT}_{ij}
= 0\,.
\end{equation}
We can take a plane-wave solution $ h^\mathrm{TT}_{ij} \propto \mathrm e^{\mathrm i\,\omega\,(z-t)} $ and obtain the linear Riemann tensor \eqref{eq:Riemann_h} as
\begin{equation}
{}^{(1)}R_{i0j0}
= -\frac{1}{2}\,\ddot h^\mathrm{TT}_{ij}
= \frac{1}{2}\,\omega^2\,\left(h^\mathrm{TT}_+\,e^+_{ij} + h^\mathrm{TT}_\times\,e^\times_{ij}\right)\,.
\end{equation}

The same conclusion can be drawn in the gauge-invariant formulation, in which the GR action \eqref{eq:action_gr} is rewritten in terms of the gauge-invariant variables as
\begin{equation}
S_\mathrm{GR}[H_{ij},\Sigma_i,\Phi,\Psi] 
= \frac{1}{2\kappa}\,\int\!\mathrm d^4x\,\left[ 
   H^{ij}\,\square H_{ij}
   + \frac{1}{2}\,\partial^j \Sigma^i\,\partial_j \Sigma_i
   - 6 \Phi\,\square\Phi
   - 4 \Phi\,\triangle (\Psi-\Phi)
  \right]\,.
\end{equation}
The eoms in vacuum are
\begin{equation}
\Box H_{ij} = 0\,,
\quad
\triangle \Sigma_i = 0\,,
\quad
3 \ddot\Phi
- \triangle \Psi
- \triangle \Phi
= 0\,,
\quad
\triangle \Phi
= 0\,.
\end{equation}
These imply $ \Sigma_i = 0 $ and $ \Phi = \Psi = 0 $ thanks to the assumed invertibility of Laplacian.
Therefore, for a plane-wave solution $ H_{ij} \propto \mathrm e^{\mathrm i\,\omega\,(z-t)} $, the Riemann tensor \eqref{eq:Riemann_gi} is
\begin{equation}
{}^{(1)}R_{i0j0}
= -\ddot H_{ij}
= \omega^2\,\left(H_+\,e^+_{ij} + H_\times\,e^\times_{ij}\right)\,.
\end{equation}

\subsection{\label{sec:mg}Generic linear massive gravity}

Next, we consider linear massive gravity (MG) specified by the action
\begin{equation}
\begin{aligned}
S_\mathrm{MG}[h_{\mu\nu}]
&
= S_\mathrm{GR}[h_{\mu\nu}]
  - \frac{m^2}{8\kappa}\,\int\!\mathrm d^4x\,\left[
     h_{\mu\nu}\,h^{\mu\nu} - (1-\epsilon)\,h^2
    \right] \\
&
= \frac{1}{4\kappa}\,\int\!\mathrm d^4x\,\left[
   -{}^{(1)}G_{\mu\nu}\,h^{\mu\nu}
   - \frac{m^2}{2}\,(h_{\mu\nu}\,h^{\mu\nu} - (1-\epsilon)\,h^2)
  \right]\,,
\end{aligned}
\label{eq:action_mg}
\end{equation}
where $ S_\mathrm{GR} $ is the GR action \eqref{eq:action_gr}, $ {}^{(1)}G_{\mu\nu} $ is the linear Einstein tensor \eqref{eq:ein}, $ m $ corresponds to the mass of spin-$ 2 $ graviton and $ \epsilon $ is a nondimensional parameter.
This is the most generic extension of linear general relativity that incorporates Lorentz-invariant mass terms.
For $ \epsilon = 0 $, the action reduces to that of the Fierz--Pauli theory \cite{Fierz:1939ix}, where the graviton is pure spin-$ 2 $, otherwise a spin-$ 0 $ ``ghost'' graviton emerges, as we shall confirm below.

In order to treat the dynamical dofs efficiently, we use the tensor (T), vector (V) and scalar (S) variables defined as \eqref{eq:h_SVT}.
The action \eqref{eq:action_mg} is decomposed as
\begin{equation}
S_\mathrm{MG}[h_{\mu\nu}]
= S_\mathrm{MG}^{(\mathrm T)}[H_{ij}]
  + S_\mathrm{MG}^{(\mathrm V)}[B_i,E_i]
  + S_\mathrm{MG}^{(\mathrm S)}[A,B,C,E]
\end{equation}
with
\begin{align}
S_\mathrm{MG}^{(\mathrm T)}[H_{ij}]
&
= \frac{1}{2\kappa}\,\int\!\mathrm d^4x\,\left[
   H_{ij}\,\square H^{ij} - m^2\,H_{ij}\,H^{ij}
  \right]\,,
\label{eq:action_mg_T} \\
S_\mathrm{MG}^{(\mathrm V)}[B_i,E_i]
&
= \frac{1}{2\kappa}\,\int\!\mathrm d^4x\,\left[
   -\frac{1}{2}\,(B_i+\dot E_i)\,\triangle (B^i+\dot E^i)
   + \frac{m^2}{2}\,\left(B_i\,B^i + E_i\,\triangle E^i\right)
  \right]\,,
\label{eq:action_mg_V} \\
S_\mathrm{MG}^{(\mathrm S)}[A,B,C,E]
&
= \frac{1}{2\kappa}\,\int\!\mathrm d^4x\,\biggl[
   -6 C\,\square C
   - 4 C\,\triangle (A-\dot B-\ddot E-C)
   - \frac{m^2}{2}\,\bigl(
      2 \epsilon\,A^2
      + B\,\triangle B
      + 6\,(3 \epsilon-2)\,C^2 \nonumber \\
& \qquad\qquad
      + 2 \epsilon\,E\,\triangle^2 E
      + 4\,(3 \epsilon-2)\,C\,\triangle E
      + 4\,(\epsilon-1)\,A\,\triangle E
      + 12\,(\epsilon-1)\,A\,C
     \bigr)
  \biggr]\,.
\label{eq:action_mg_S}
\end{align}
Contrary to GR, these vector and scalar actions cannot be solely expressed with the gauge-invariant variables due to the lack of the gauge symmetries.
Thus, in order to calculate the Riemann tensor \eqref{eq:Riemann_gi}, we have to manipulate the original variables.

The eom for the tensor variable $ H_{ij} $ is obtained from \eqref{eq:action_mg_T} as
\begin{equation}
\square H_{ij}
- m^2\,H_{ij}
= 0\,.
\end{equation}
The eom for the gauge-invariant variable $ \Sigma_i $ does not directly derive, but the eoms for $ B_i $ and $ E_i $ from \eqref{eq:action_mg_V} are
\begin{equation}
-\triangle (B_i + \dot E_i)
+ m^2\,B_i
= 0\,,
\quad
\dot B_i + \ddot E_i
+ m^2\,E_i
= 0\,.
\label{eq:eom_mg_V}
\end{equation}
Combining these, we find that $ \Sigma_i \equiv B_i + \dot E_i $ obeys
\begin{equation}
\square \Sigma_i - m^2\,\Sigma_i
= 0\,.
\end{equation}
$ \Sigma_i $ embodies all the dynamical vector-type dofs as $ B_i $ and $ E_i $ are dependent upon $ \Sigma_i $ via the nondynamical relations
\begin{equation}
B_i
= m^{-2}\,\triangle \Sigma_i\,,
\quad
E_i
= -m^{-2}\,\dot\Sigma_i
\end{equation}
from \eqref{eq:eom_mg_V}.
The most complicated is to find the governing equations for the gauge-invariant scalars $ \Psi $ and $ \Phi $.
The scalar eoms from variations of \eqref{eq:action_mg_S} with respect to $ A, B, C, E $ are, respectively,
\begin{equation}
\begin{aligned}
&
2 \triangle C
+ m^2\,(\epsilon A + (\epsilon-1)\,\triangle E + 3\,(\epsilon-1)\,C)
= 0\,, \\
&
\triangle\left[4 \dot C + m^2\,B\right]
= 0\,, \\
&
6 \square C
+ 2 \triangle (A - \dot B - \ddot E)
- 4 \triangle C
+ m^2\,(3\,(3\epsilon-2)\,C + (3\epsilon-2)\,\triangle E + 3\,(\epsilon-1)\,A)
= 0\,, \\
&
\triangle\left[
 -2 \ddot C
 + m^2\,(\epsilon \triangle E + (3\epsilon-2)\,C + (\epsilon-1)\,A)
\right]
= 0\,.
\end{aligned}
\label{eq:eom_mg_S}
\end{equation}
Gathering these equations, it is found that the following two combinations
\begin{equation}
W
\equiv
  A - \dot B - \ddot E - C\,,
\quad
h
\equiv
  2 A + 6 C + 2 \triangle E
\end{equation}
satisfy Klein--Gordon equations
\begin{equation}
\square W - m^2\,W
= 0\,,
\quad
\epsilon\,\square h - \frac{3-4\epsilon}{2}\,m^2\,h
= 0\,.
\label{eq:eom_mg_Wh}
\end{equation}
Observe that $ W = \Psi - \Phi $ and $ h = \eta^{\mu\nu}\,h_{\mu\nu} $\,.
The second equation implies that, if the Fierz--Pauli tuning $ \epsilon = 0 $ is realized, then the four-dimensional trace $ h $ is constrained to vanish.
We assume $ \epsilon \neq 0 $ and define $ m_0^2 \equiv \frac{3-4\epsilon}{2 \epsilon}\,m^2 $ as the mass of $ h $.
These $ W $ and $ h $ are the only dynamical scalar-type dofs.
Indeed, the set of equations~\eqref{eq:eom_mg_S} can be solved for $ A, B, C, E $ in terms of $ W $ and $ h $ as
\begin{equation}
\begin{aligned}
A
&
= \frac{2}{3 m^4}\,\triangle^2 W
  + \left(\frac{1-\epsilon}{2} - \frac{\epsilon}{3 m^2}\,\triangle\right)\,h\,, \\
B
&
= \frac{4}{3 m^4}\,\triangle \dot W - \frac{2 \epsilon}{3 m^2}\,\dot h\,, \\
C
&
= -\frac{1}{3 m^2}\,\triangle W + \frac{\epsilon}{6}\,h\,, \\
E
&
= \frac{1}{m^2}\,W - \frac{2}{3 m^4}\,\triangle W + \frac{\epsilon}{3 m^2}\,h\,.
\end{aligned}
\end{equation}
Moreover, the gauge-invariant variables are expressed as
\begin{equation}
\Psi
= A - \dot B - \ddot E
= W - \frac{1}{3 m^2}\,\triangle W + \frac{\epsilon}{6}\,h\,,
\quad
\Phi
= C
= -\frac{1}{3 m^2}\,\triangle W + \frac{\epsilon}{6}\,h\,,
\end{equation}
where we have used \eqref{eq:eom_mg_Wh} to eliminate $ \ddot W $ and $ \ddot h $.
Using these relationships, the scalar part of the linear Riemann tensor is written as
\begin{equation}
\begin{aligned}
{}^{(1)}R^{(\mathrm S)}_{i0j0}
&
\equiv
  \partial_i \partial_j \Psi
  - \delta_{ij}\,\ddot\Phi \\
&
= 3 \partial_i \partial_j W
  - \frac{1}{3 m^2}\,
    \triangle \left(\partial_i \partial_j W - \delta_{ij}\,\ddot W\right)
  + \frac{\epsilon}{6}\,
    \left(\partial_i \partial_j h - \delta_{ij}\,\ddot h\right)\,.
\end{aligned}
\label{eq:Riemann_SVT}
\end{equation}

Having found that the variables $ H_{ij} $\,, $ \Sigma_i $\,, $ W $ and $ h $ all obey Klein--Gordon-type equations, we are allowed to consider plane-wave solutions propagating along the $ z $ direction,
\begin{equation}
H_{ij}
\propto
  \mathrm e^{\mathrm i\,(k_H z - \omega_H t)}\,,
\quad
\Sigma_i
\propto
  \mathrm e^{\mathrm i\,(k_\Sigma z - \omega_\Sigma t)}\,,
\quad
W
\propto
  \mathrm e^{\mathrm i\,(k_W z - \omega_W t)}\,,
\quad  
h
\propto
  \mathrm e^{\mathrm i\,(k_h z - \omega_h t)}
\end{equation}
with
\begin{equation}
k_I
\equiv
  \sqrt{\omega_I^2 - m^2}
\quad(I = H, \Sigma, W)\,,
\quad
k_h
\equiv
  \sqrt{\omega_h^2 - m_0^2}\,.
\end{equation}
Then the Riemann tensor is calculated as
\begin{equation}
\begin{aligned}
{}^{(1)}R_{i0j0}
&
= \omega_H^2\,(H_+\,e^+_{ij} + H_\times\,e^\times_{ij})
  - \frac{1}{2}\,\sqrt{\omega_\Sigma^2-m^2}\,\omega_\Sigma\,
    \left(\Sigma_x\,e^x_{ij} + \Sigma_y\,e^y_{ij}\right) \\
& \quad
  - \frac{1}{3m^2}\,\triangle W\,
    \left(
     \omega_W^2\,e^\mathrm B_{ij}
     - \sqrt 2\,m^2\,e^\mathrm L_{ij}
    \right)
  + \frac{\epsilon}{6}\,h\,
    \left(
     \omega_h^2\,e^\mathrm B_{ij}
     + \frac{m_0^2}{\sqrt 2}\,e^\mathrm L_{ij}
    \right)\,.
\end{aligned}
\label{eq:Riemann_mg}
\end{equation}
This expression tells us that the separated six variables provide different polarizations.
In particular, the information carried by the two scalar variables is distinctive.
$ W $, the helicity-$ 0 $ mode of the spin-$ 2 $ graviton, can be split into the transverse, or  ``breathing'' ($ \mathrm B $), and longitudinal ($ \mathrm L $) polarizations based on its different dependences on the frequency:
the former is proportional to $ \omega_W^2 $ while the latter is $ m^2 $.
Similarly, the spin-$ 0 $ graviton $ h $, which only exists if $ \epsilon \neq 0 $, can be decomposed into the transverse and longitudinal polarizations.
Thus, if the amplitudes of each polarization are separately measured in future gravitational-wave experiments, the longitudinal modes will provide a direct measure of the masses of the spin-$ 2 $ and spin-$ 0 $ gravitons.
We will come back to this issue in Sec.~\ref{sec:obs}.

Finally, let us mention the effectively massless case $ m^2 \ll \omega_I^2 $.
Resembling the discontinuity in the bending angle of light \cite{Iwasaki:1971uz,vanDam:1970vg,Zakharov:1970cc}, taking the continuous limit does not recover the set of polarizations expected in GR as the vector and the transverse scalar polarizations would remain.

\subsection{\label{sec:hcg}Generic higher-curvature gravity}

Next we consider a class of extended theories of gravity whose full action is of the form
\begin{equation}
S
= \frac{1}{2\kappa}\,
  \int\!\mathrm d^4x\,\sqrt{-g}\,f(R^\mu{}_{\nu\rho\sigma},g_{\mu\nu})\,,
\end{equation}
where $ f $ contains terms non-linear in the Riemann curvature.
We begin with discussing perturbative degrees of freedom in this higher-curvature gravity (HCG) in the Minkowski background.
If the Lagrangian $ f $ consists only of terms that have smooth behavior around $ R_{\mu\nu\rho\sigma} = 0 $, its expansion in curvature tensors up to the quadratic order can be arranged as
\begin{equation}
f
= \chi\,R
  - \alpha\,C_{\mu\nu\rho\sigma}\,C^{\mu\nu\rho\sigma}
  + \beta\,R^2
  + \gamma\,\left(
     R_{\mu\nu\rho\sigma}\,R^{\mu\nu\rho\sigma} - 4\,R_{\mu\nu}\,R^{\mu\nu} + R^2
    \right)
  + \mathcal O(R_{\mu\nu\rho\sigma})^3\,,
\label{eq:L_hcg}
\end{equation}
where $ \chi $, $ \alpha $, $ \beta $ and $ \gamma $ are constants and $ C_{\mu\nu\rho\sigma} $ is the Weyl curvature tensor.
The combination in the parentheses, so-called Gauss--Bonnet invariant, is topological in four dimensions and can be discarded in the action integral.
Then we find that the generic higher-curvature action expanded up to the second order in the metric perturbation $ h_{\mu\nu} $ 
\begin{equation}
S_\mathrm{HCG}[h_{\mu\nu}]
= \frac{1}{2\kappa}\,\int\!\mathrm d^4x\,\left(
   -\frac{\chi}{2}\,{}^{(1)}G_{\mu\nu}\,h^{\mu\nu}
   - \alpha\,{}^{(1)}C_{\mu\nu\rho\sigma}\,{}^{(1)}C^{\mu\nu\rho\sigma}
   + \beta\,{}^{(1)}R^2
  \right)\,,
\label{eq:action_hcg}
\end{equation}
where $ {}^{(1)}C_{\mu\nu\rho\sigma} $ and $ {}^{(1)}R $ are the linear perturbation of the Weyl tensor and Ricci scalar, respectively, whose expressions are presented in Appendix~\ref{sec:pert}.
When $ \chi = 0 $, the theory cannot be seen as GR with corrections and, moreover, as we will see in Appendix~\ref{sec:noGR}, there arise instabilities in the tensor and scalar parts.
So, hereafter we assume $ \chi \neq 0 $.

Below, we take two different approaches to analyze the GWs in generic HCG.

\subsubsection{\label{sec:hcg_bi}Massive-bigravity approach}

As first shown by Stelle \cite{Stelle:1977ry}, there is an equivalence of the action \eqref{eq:action_hcg} to GR ``minus'' massive gravity,
\begin{equation}
\begin{aligned}
S[\phi_{\mu\nu},\tilde\phi_{\mu\nu}]
&
= \chi\,S_\mathrm{GR}[\phi_{\mu\nu}] - \chi\,S_\mathrm{MG}[\tilde\phi_{\mu\nu}] \\
&
= \frac{\chi}{4\kappa}\,\int\!\mathrm d^4x\,\left[
   -{}^{(1)}G_{\mu\nu}[\phi]\,\phi^{\mu\nu}
   + {}^{(1)}G_{\mu\nu}[\tilde\phi]\,\tilde\phi^{\mu\nu}
   + \frac{m^2}{2}\,
     \left(\tilde\phi_{\mu\nu}\,\tilde\phi^{\mu\nu} - (1-\epsilon)\,\tilde\phi^2\right)
  \right]\,,
\end{aligned}
\label{eq:action_bi}
\end{equation}
where we introduced $ m^2 \equiv \chi/(2\alpha) $ and $ \epsilon = 9\beta/(2\alpha+12\beta) $, see Appendix~\ref{sec:dec} for the derivation extended to Einstein manifolds \cite{Niiyama:2019fvf}.
Clearly, we need to assume $ \alpha \neq 0 $.
The case with $ \alpha = 0 $ can be treated in a similar manner by means of a conformal transformation, after which the theory takes a form of a scalar--tensor theory;
See \cite{Moretti:2019yhs} for example.

In this formalism, the original metric perturbation is given by
\begin{equation}
h_{\mu\nu}
= \phi_{\mu\nu}
  + \tilde\phi_{\mu\nu}\,,
\end{equation}
and, as seen from the structure of the action \eqref{eq:action_bi}, the dynamical contents in this theory are $ \phi_{\mu\nu} $ as a massless spin-$ 2 $ field and $ \tilde\phi_{\mu\nu} $ as a mixture of a massive spin-$ 2 $ and a spin-$ 0 $ fields.
It is worth mentioning that, while $ \phi_{\mu\nu} $ field has the same gauge symmetry as GR, $ \tilde\phi_{\mu\nu} $ is not subject to gauge transformations.

For the massless spin-$ 2 $, we can choose the TT gauge as in GR treated in Sec.~\ref{sec:GR}.
The only dynamical dof is $ \phi^\mathrm{TT}_{ij} = 2 H_{ij} $ with its eom being $ \square H_{ij} = 0 $.
For the massive spin-$ 2 $, the analysis is completely parallel to the case of massive gravity treated in Sec.~\ref{sec:mg}.
We decompose $ \tilde\phi_{\mu\nu} $ as
\begin{equation}
\tilde\phi_{00}
= -2 \tilde A\,,
\quad
\tilde\phi_{0i}
= -\partial_i \tilde B - \tilde B_i\,,
\quad
\tilde\phi_{ij}
= 2 \tilde C\,\delta_{ij}
  + 2 \partial_i \partial_j \tilde E
  + 2 \partial_{(i} \tilde E_{j)}
  + 2 \tilde H_{ij}
\end{equation}
and define two scalar variables
\begin{equation}
\tilde W
\equiv
  \tilde A - \dot{\tilde B} - \ddot{\tilde E} - \tilde C\,,
\quad
\tilde\phi
\equiv
  \tilde\phi_\mu{}^\mu
  = 2 \tilde A + 6 \tilde C + 2 \triangle \tilde E\,.
\end{equation}
Then we find the eoms for the dynamical variables
\begin{equation}
\left(\square-m^2\right)\,\tilde H_{ij}
= 0\,,
\quad
\left(\square-m^2\right)\,\tilde\Sigma_i
= 0\,,
\quad
\left(\square-m^2\right)\,\tilde W
= 0\,,
\quad
\left(\frac{6 \beta}{\chi}\,\square-1\right)\,\tilde\phi
= 0\,.
\end{equation}
In this case, the Fierz--Pauli tuning $ \epsilon = 0 $ is realized and the spin-$ 0 $ mode $ \tilde\phi $ is required to vanish when $ \beta = 0 $, otherwise it acquires a finite mass $ m_0^2 \equiv \chi/(6\beta) $.
Considering plane wave solutions
\begin{equation}
H_{ij}
\propto
  \mathrm e^{\mathrm i\,\omega_H\,(z - t)}\,,
\quad
\tilde H_{ij}
\propto
  \mathrm e^{\mathrm i\,(k_{\tilde H}\,z - \omega_{\tilde H}\,t)}\,,
\quad
\tilde\Sigma_i
\propto
  \mathrm e^{\mathrm i\,(k_{\tilde\Sigma}\,z - \omega_{\tilde\Sigma}\,t)}\,,
\quad
\tilde W
\propto
  \mathrm e^{\mathrm i\,(k_{\tilde W}\,z - \omega_{\tilde W}\,t)}\,,
\quad
\tilde\phi
\propto
  \mathrm e^{\mathrm i\,(k_{\tilde\phi}\,z - \omega_{\tilde\phi}\,t)}
\end{equation}
with dispersion relations
\begin{equation}
k_I
=
\begin{cases}
\sqrt{\omega_I^2 - m^2}
& (I = \tilde H, \tilde\Sigma, \tilde W) \\
\sqrt{\omega_I^2 - m_0^2}
& (I = \tilde\phi)
\end{cases}\,,
\end{equation}
we get the following expression for the Riemann tensor:
\begin{equation}
\begin{aligned}
{}^{(1)}R_{i0j0}
&
= \sum_{\lambda=+,\times}
  \left[
   \omega_H^2\,H_\lambda + \omega_{\tilde H}^2\,\tilde H_\lambda
  \right]\,
  e^\lambda_{ij}
  - \frac{1}{2}\,\sqrt{\omega_{\tilde\Sigma}^2-m^2}\,\omega_{\tilde\Sigma}\,
    \sum_{p=x,y} \tilde\Sigma_p\,e^p_{ij} \\
& \quad
  - \frac{1}{3 m^2}\,\triangle \tilde W\,
    \left(
     \omega_{\tilde W}^2\,e^\mathrm B_{ij}
     - \sqrt 2\,m^2\,e^\mathrm L_{ij}
    \right)
  + \frac{\epsilon}{6}\,\tilde\phi\,
    \left(
     \omega_{\tilde\phi}^2\,e^\mathrm B_{ij}
     + \frac{m_0^2}{\sqrt 2}\,e^\mathrm L_{ij}
    \right)\,.
\end{aligned}
\label{eq:Riemann_hcg}
\end{equation}

To summarize, we have established that the gravitational-wave polarizations in generic HCG with $ \alpha \neq 0 $ is a sum of those in GR and in MG, with a difference that the mass parameters $ m $ and $ m_0 $ in this case are given by the expansion coefficients $ \chi $, $ \alpha $ and $ \beta $ inherent in the nonlinear Lagrangian $ f $.

\subsubsection{\label{sec:hcg_gi}Gauge-invariant approach}

Next, we take a distinct approach starting with introduction of the gauge-invariant variables.
We first assume that $ \chi \neq 0 $ as in the previous section, but $ \alpha \neq 0 $ is not mandatory.
We will discuss some special cases in the main text and in Appendix~\ref{sec:noGR}.

Using the result presented in Appendix~\ref{sec:pert}, the action \eqref{eq:action_hcg} can be rewritten in terms of the gauge-invariant variables as
\begin{equation}
S_\mathrm{HCG}[h_{\mu\nu}] 
= S_\mathrm{HCG}^\mathrm{(T)}[H_{ij}]
  + S_\mathrm{HCG}^\mathrm{(V)}[\Sigma_i]
  + S_\mathrm{HCG}^\mathrm{(S)}[\Phi,\Psi]\,,
\end{equation}
where each part is
\begin{align}
S_\mathrm{HCG}^{(\mathrm T)}[H_{ij}]
&
= \frac{1}{2\bar\kappa}\,\int\!\mathrm d^4x\,\left[
   H_{ij}\,\square H^{ij} - 2 \bar\alpha\,\square H_{ij}\,\square H^{ij}
  \right]\,,
\label{eq:action_hcg_t} \\
S_\mathrm{HCG}^{(\mathrm V)}[\Sigma_i]
&
= \frac{1}{2\bar\kappa}\,\int\!\mathrm d^4x\,\left[
   \frac{1}{2}\,\partial_j \Sigma_i\,\partial^j \Sigma^i
   - \bar\alpha\,\left(
      \partial_j \dot\Sigma_i\,\partial^j \dot\Sigma^i
      - \triangle \Sigma_i\,\triangle \Sigma^i
     \right)
  \right]\,,
\label{eq:action_hcg_v} \\
S_\mathrm{HCG}^{(\mathrm S)}[\Phi,\Psi]
&
= \frac{1}{2\bar\kappa}\,\int\!\mathrm d^4x\,\left[
   -6 \Phi\,\square \Phi - 4 \Phi\,\triangle (\Psi-\Phi)
   - \frac{4}{3}\,\bar\alpha\,(\triangle \Psi - \triangle \Phi)^2
   + \bar\beta\,{}^{(1)}R^2
  \right]
\label{eq:action_hcg_s}
\end{align}
with $ {}^{(1)}R = -6 \square\Phi - 2 \triangle (\Psi-\Phi) $, $ \bar\kappa \equiv \kappa/\chi $, $ \bar\alpha \equiv \alpha/\chi $ and $ \bar\beta \equiv \beta/\chi $.
As seen in the above expressions, when $ \alpha = 0 $, the tensor and vector parts reduce to the ones in GR.
Conversely, the tensor and vector parts are modified in comparison to GR by the presence of the Weyl-squared term but unaffected by the Ricci-squared term in the expansion of the Lagrangian \eqref{eq:L_hcg}.
On the other hand, the scalar part is affected by both the Weyl-squared and Ricci-squared terms.
In the following, we proceed to the analyses for each part.

First, the tensor part is only affected by the presence of the Weyl term as seen in \eqref{eq:action_hcg_t}.
The tensor eom is
\begin{equation}
\square H_{ij}
- 2 \bar\alpha\,\square^2 H_{ij}
= 0\,.
\end{equation}
When $ \alpha = 0 $, the tensor eom reduces to that of GR, and the same result for the polarizations is obtained.
When $ \alpha \neq 0 $, we find that the above eom admits two independent solutions $ H_{ij} = \phi_{ij} $ and $ H_{ij} = \tilde\phi_{ij} $ which respectively satisfy
\begin{equation}
\square \phi_{ij}
= 0\,,
\quad
\square \tilde\phi_{ij} - m^2\,\tilde\phi_{ij}
= 0\,,
\end{equation}
where, as before, $ m^2 = 1/(2\bar\alpha) $.
Hence the general solution for $ H_{ij} $ is
\begin{equation}
H_{ij}
= \phi_{ij} + \tilde\phi_{ij}\,.
\end{equation}
Considering plane-wave solutions propagating in the $ z $ direction,
\begin{equation}
\phi_{ij}
\propto
  \mathrm e^{\mathrm i\,\omega_\phi\,(z-t)} \,,
\quad
\tilde\phi_{ij}
\propto
  \mathrm e^{\mathrm i\,(k_{\tilde\phi}\,z-\omega_{\tilde\phi}\,t)}\,,
\end{equation}
where $ k_{\tilde\phi} = \sqrt{\omega_{\tilde\phi}^2-m^2} $, and substituting these into \eqref{eq:Riemann_gi}, we obtain the tensor part of the Riemann tensor as
\begin{equation}
{}^{(1)}R_{i0j0}^{(\mathrm T)}
= \omega_\phi^2\,\left(\phi_+\,e^+_{ij} + \phi_\times\,e^\times_{ij}\right)
  + \omega_{\tilde\phi}^2\,
    \left(\tilde\phi_+\,e^+_{ij} + \tilde\phi_\times\,e^\times_{ij}\right)\,.
\end{equation}
Identifying $ \phi_\lambda $ with $ H_\lambda $ and $ \tilde\phi_\lambda $ with $ \tilde H_\lambda $\,, we reproduce the tensor part of \eqref{eq:Riemann_hcg}.

Next, as for the vector part, the eom from the action \eqref{eq:action_hcg_v} is
\begin{equation}
\left(1 - 2 \bar\alpha\,\square\right)\,\triangle \Sigma_i = 0\,.
\end{equation}
When $ \alpha = 0 $, $ \Sigma_i = 0 $ as expected.
When $ \alpha \neq 0 $, the eom admits a plane-wave solution
\begin{equation}
\Sigma_i
\propto
  \mathrm e^{\mathrm i\,(k_\Sigma\,z-\omega_\Sigma\,t)}
\end{equation}
with $ k_\Sigma = \sqrt{\omega_\Sigma^2-m^2} $.
For this, the vector part of the Riemann tensor \eqref{eq:Riemann_gi} is
\begin{equation}
{}^{(1)}R_{i0j0}^{(\mathrm V)}
= -\frac{1}{2}\,\sqrt{\omega_\Sigma^2-m^2}\,\omega_\Sigma\,
  \left(\Sigma_x\,e^x_{ij} + \Sigma_y\,e^y_{ij}\right)\,,
\end{equation}
which is identical with the vector part of \eqref{eq:Riemann_hcg}.

Finally, we analyze the scalar part.
As for scalar, counting of the number of dofs is a nontrivial task since the action \eqref{eq:action_hcg_s} contains second-order time derivatives nonlinearly in the Ricci-squared term.
Let us first introduce a variable $ \Theta = \frac{2}{3}\,\triangle (\Phi - \Psi) $ to simplify the scalar action as
\begin{equation}
S_\mathrm{HCG}^{(\mathrm S)}[\Phi,\Theta]
= \frac{1}{2\bar\kappa}\,\int\!\mathrm d^4x\,\left[
   -6 \Phi\,\square\Phi
   + 6 \Phi\,\Theta
   - 3 \bar\alpha\,\Theta^2
   + \bar\beta\,{}^{(1)}R^2
  \right]\,.
\end{equation}
To have an equivalent action with only derivatives lower than or equal to second order, we replace the Ricci scalar $ {}^{(1)}R = -6 \square\Phi + 3 \Theta $ with an auxiliary variable $ \Xi $ introducing a Lagrange multiplier $ \lambda $ as
\begin{equation}
S_\mathrm{HCG}^{(\mathrm S)}[\Phi,\Theta,\Xi,\lambda]
= \frac{1}{2\bar\kappa}\,\int\!\mathrm d^4x\,\left[
   -6 \Phi\,\square\Phi
   + 6 \Phi\,\Theta
   - 3 \bar\alpha\,\Theta^2
   + \bar\beta\,\Xi^2
   + \lambda\,({}^{(1)}R - \Xi)
  \right]\,.
\end{equation}
The variation of the above action with respect to $ \Xi $ gives a constraint $ \lambda = 2 \bar\beta\,\Xi $, which can be used to eliminate $ \lambda $ as
\begin{equation}
S_\mathrm{HCG}^{(\mathrm S)}[\Phi,\Theta,\Xi]
= \frac{1}{2\bar\kappa}\,\int\!\mathrm d^4x\,\left[
   -6 \Phi\,\square\Phi
   + 6 \Phi\,\Theta
   - 3 \bar\alpha\,\Theta^2
   - \bar\beta\,\Xi^2
   - 12 \bar\beta\,\Xi\,\square\Phi
   + 6 \bar\beta\,\Xi\,\Theta
  \right]\,.
\end{equation}
Now, after some manipulation, we obtain three independent equations from the variations of the above action with respect to each variable,
\begin{align}
&
\Theta
- 2 \bar\alpha\,\square\Theta
= 0\,,
\label{eq:eom_Theta} \\
&
\bar\beta\,\Xi
- 6 \bar\beta^2\,\square\Xi
= 0\,,
\label{eq:eom_Xi} \\
&
\Phi
= \bar\alpha\,\Theta
  - \bar\beta\,\Xi\,.
\label{eq:const_Phi}
\end{align}
The eoms \eqref{eq:eom_Theta} and \eqref{eq:eom_Xi} imply that $ \Theta $ and $ \Xi $ are independent dofs with masses $ m^2 = 1/(2\bar\alpha) $ and $ m_0^2 = 1/(6\bar\beta) $, respectively.
The algebraic constraint \eqref{eq:const_Phi} indicates that when $ \bar\alpha \neq 0 $ ($ \bar\beta \neq 0 $), $ \Theta $ ($ \Xi $) can be eliminated.
When both $ \bar\alpha $ and $ \bar\beta $ are nonzero, we can solve \eqref{eq:eom_Theta} and \eqref{eq:eom_Xi} for $ \Theta $ and $ \Xi $ and obtain $ \Phi $ via \eqref{eq:const_Phi}.
We then have the other gauge-invariant variable
\begin{equation}
\Psi
= \bar\alpha\,\Theta - \frac{3}{2}\,\triangle^{-1} \Theta - \bar\beta\,\Xi\,.
\end{equation}
Assuming plane-wave solutions
\begin{equation}
\Theta
\propto
  \mathrm e^{\mathrm i\,\left(k_\Theta\,z - \omega_\Theta\,t\right)}\,,
\quad
\Xi
\propto
  \mathrm e^{\mathrm i\,\left(k_\Xi\,z - \omega_\Xi\,t\right)}
\end{equation}
with $ k_\Theta = \sqrt{\omega_\Theta^2-m^2} $ and $ k_\Xi = \sqrt{\omega_\Xi^2-m_0^2} $, we arrive at the expression for the scalar part of the linear Riemann tensor \eqref{eq:Riemann_gi}
\begin{equation}
{}^{(1)}R_{i0j0}^{(\mathrm S)}
= \bar\alpha\,\Theta\,
  \left(\omega_\Theta^2\,e^\mathrm B_{ij} - \sqrt 2\,m^2\,e^\mathrm L_{ij}\right)
  - \bar\beta\,\Xi\,
    \left(\omega_\Xi^2\,e^\mathrm B_{ij} + \frac{m_0^2}{\sqrt 2}\,e^\mathrm L_{ij}\right)\,.
\end{equation}
This is identical with the scalar part of \eqref{eq:Riemann_hcg} if we identify as
\begin{equation}
\bar\alpha\,\Theta
= -\frac{1}{3 m^2}\,\triangle \tilde W\,,
\quad
\bar\beta\,\Xi
= -\frac{\epsilon}{6}\,\tilde\phi\,.
\end{equation}

Our final task here is to investigate special cases with $ \alpha = 0 $ or $ \beta = 0 $.
When $ \alpha = 0 $, the eom \eqref{eq:eom_Theta} reduces to a constraint $ \Theta = 0 $, which implies $ \Psi = \Phi $.
Equation~\eqref{eq:const_Phi} reduces to $ \Phi = -\bar\beta\,\Xi $, so the variable $ \Phi $ represents the spin-$ 0 $ dof and obeys the eom
\begin{equation}
\square \Phi - m_0^2\,\Phi
= 0\,.
\end{equation}
Therefore, recalling $ \Psi = \Phi $, we find the scalar-type polarization in \eqref{eq:Riemann_gi} in this case as
\begin{equation}
{}^{(1)}R^{(\mathrm S)}_{i0j0}
= \Phi\,
  \left(
   \omega_\Phi^2\,e^\mathrm B_{ij}
   + \sqrt 2\,m_0^2\,e^\mathrm L_{ij}
  \right)\,.
\end{equation}
Also, the vector and tensor perturbations give the same result as in GR.
This agrees with the result of Moretti \textit{et al.}~\cite{Moretti:2019yhs}.

When $ \beta = 0 $, the constraint \eqref{eq:const_Phi} reduces to $ \Phi = \bar\alpha\,\Theta $\,, which implies that
\begin{equation}
\Psi
= \Phi - \frac{3}{2\bar\alpha}\,\triangle^{-1} \Phi
\label{eq:const_Psi}
\end{equation}
and the eom for $ \Phi $ is
\begin{equation}
\square \Phi - m^2\,\Phi
= 0\,.
\end{equation}
It is clear that $ \Phi $ in this case is the helicity-$ 0 $ component of the massive spin-$ 2 $.
Recalling its relation to $ \Psi $ as given by \eqref{eq:const_Psi}, the scalar-type polarization in \eqref{eq:Riemann_gi} is calculated as
\begin{equation}
{}^{(1)}R^{(\mathrm S)}_{i0j0}
= \Phi\,
  \left(
   \omega_\Phi^2\,e^\mathrm B_{ij}
   - \frac{m^2}{\sqrt 2}\,e^\mathrm L_{ij}
  \right)\,.
\end{equation}

\section{\label{sec:obs}Determining theory parameters by observations}

In this section, we provide a brief discussion of how one could determine the theory parameters, namely $ m $ and $ \epsilon $ in MG or $ \alpha $ and $ \beta $ in HCG.
For brevity, below we collectively call the theory parameters ``masses.''
We consider interferometers and pulsar timing arrays (PTAs) as GW measurement instruments.

To discuss the ability of GW detectors for determining the masses via polarization measurements, it is necessary to take into account detector's response to each polarization in a GW propagating from the direction $ (\theta,\phi) $ with the polarization angle $ \psi $, which is represented by the antenna pattern functions $ F_\alpha(\theta,\phi,\psi) $ ($ \alpha = +,\times,x,y,\mathrm B,\mathrm L $) summarized in Appendix~\ref{sec:F}.
The whole signal takes the form
\begin{equation}
S(t)
= \sum_\alpha F_\alpha(\theta,\phi,\psi)\,A_\alpha(t)\,,
\end{equation}
where $ A_\alpha $ is the waveform of each polarization being proportional to the coefficient of the polarization basis $ e^\alpha_{ij} $ appearing in the Riemann tensor.

We note that different spin components have different velocity, so it would be plausible to treat each spin component separately if one is interested in the case of a short-duration source like a burst from a black-hole merger.
Thus we first concentrate on the spin-$ 0 $ GW alone, denoting its velocity as $ v $.
A spin-$ 0 $ GW represented as a monochromatic plane-wave with frequency $ \omega $ gives a signal of the form
\begin{equation}
S(t)
= F_\mathrm B\,A_\mathrm B(t) + F_\mathrm L\,A_\mathrm L(t)\,.
\end{equation}
From \eqref{eq:Riemann_mg} or \eqref{eq:Riemann_hcg} we see that there is a relationship between the waveforms
\begin{equation}
\frac{A_\mathrm L(t)}{A_\mathrm B(t)}
= \frac{m_0^2}{\omega^2}\,.
\end{equation}
Suppose that two detectors ($ D = 1,2 $) respond to a single spin-$ 0 $ gravitational wave.
In this case, the signals to be detected by the two detectors are
\begin{equation}
\left\{
\begin{aligned}
S^1(t)
&
= F^1_\mathrm B\,A_\mathrm B(t) + F^1_\mathrm L\,A_\mathrm L(t)\,, \\
S^2(t+\Delta t)
&
= F^2_\mathrm B\,A_\mathrm B(t) + F^2_\mathrm L\,A_\mathrm L(t)\,,
\end{aligned}
\right.
\end{equation}
where $ \Delta t = L/v $ is the time delay between the arrivals at the two detectors separated by $ L $ along the propagation direction.
If the coefficient matrix
\begin{equation}
\mathcal F
\equiv
\begin{pmatrix}
F^1_\mathrm B & F^1_\mathrm L \\
F^2_\mathrm B & F^2_\mathrm L
\end{pmatrix}
\end{equation}
is invertible, it is possible to solve for $ A_\mathrm B $ and $ A_\mathrm L $ as
\begin{equation}
\begin{pmatrix}
A_\mathrm B(t) \\
A_\mathrm L(t)
\end{pmatrix}
=
\mathcal F^{-1}
\begin{pmatrix}
S^1(t) \\
S^2(t+\Delta t)
\end{pmatrix}\,.
\end{equation}
From this, the ratio of the waveforms $ \mathcal R \equiv A_\mathrm L/A_\mathrm B $ is obtained in terms of observable signals and, therefore, the spin-$ 0 $ mass can be determined as
\begin{equation}
m_0^2
= \omega^2\,\mathcal R\,.
\end{equation}

The above argument relies upon the invertibility of the matrix $ \mathcal F $.
Unfortunately, LIGO-like interferometers have degenerate antenna pattern functions
\begin{equation}
F_\mathrm B
= -\frac{1}{2}\,\sin^2\theta\,\cos 2\phi\,,
\quad
F_\mathrm L
= \frac{1}{\sqrt 2}\,\sin^2\theta\,\cos 2\phi\,,
\end{equation}
so it is not possible to use the above method.
On the other hand, PTAs have antenna functions \cite{Yunes:2013dva,Lee:2010cg,Qin:2020hfy}
\begin{equation}
F_\mathrm B
= \frac{1}{2}\,\frac{\sin^2\theta}{1+v\,\cos\theta}\,,
\quad
F_\mathrm L
= \frac{1}{\sqrt 2}\,\frac{\cos^2\theta}{1+v\,\cos\theta}\,,
\end{equation}
which are nondegenerate if the two detectors (pulsars) have different orientations with respect to the GW, $ \theta_1 \neq \theta_2 $\,, so they would allow the determination of the spin-$ 0 $ mass.
A straightforward extension to a setup with more detectors (pulsars) enables us to decompose a spin-$ 2 $ GW into polarization modes and, in principle, determine the spin-$ 2 $ mass $ m $ as well.
Such multidetector measurement would also be needed to analyze GW backgrounds.

In an ideal case in which we know, or can predict, the spectral form of the spin-$ 0 $ GWs $ A_\alpha(\omega,t) $, i.e., its frequency dependence, we might be able to determine the mass even with a single detector.
Consider two measurements of a GW at two different frequencies $ \omega_1 $ and $ \omega_2 $
\begin{equation}
\left\{
\begin{aligned}
S(\omega_1)
&
= F_\mathrm B\,A_\mathrm B(\omega_1) + F_\mathrm L\,A_\mathrm L(\omega_1)\,, \\
S(\omega_2)
&
= F_\mathrm B\,A_\mathrm B(\omega_2) + F_\mathrm L\,A_\mathrm L(\omega_2)\,,
\end{aligned}
\right.
\end{equation}
where we omitted $ t $.
Substituting the relationship $ A_\mathrm L(\omega)/A_\mathrm B(\omega) = m_0^2/\omega^2 $ and solving for $ m_0 $\,, we obtain
\begin{equation}
m_0^2
= \frac{F_\mathrm B}{F_\mathrm L}\,
  \frac{A_\mathrm B(\omega_1)/A_\mathrm B(\omega_2) - S(\omega_1)/S(\omega_2)}
       {\omega_2^{-2}\,S(\omega_1)/S(\omega_2) - \omega_1^{-2}\,A_\mathrm B(\omega_1)/A_\mathrm B(\omega_2)}
\end{equation}
Since the ratio $ A_\mathrm B(\omega_1)/A_\mathrm B(\omega_2) $ is assumed to be known, $ m_0^2 $ can be measured.
This method can also be extended to the determination of the spin-$ 2 $ mass $ m $.

We emphasize the merit of our methods that the determination of the mass parameter $ m_0 $ (or $ \beta $) can be carried out without measuring the velocity of the spin-$ 0 $ GW, whether absolute or relative to other signals, or analyzing the details of the waveforms.
Naturally, this advantage is also enjoyed in the relevant analyses on the spin-$ 2 $ part.

\section{\label{sec:concl}Conclusion}

In this paper, we studied gravitational-wave polarizations in generic linear massive gravity and generic higher-curvature gravity in the Minkowski background.
We defined and analyzed the GW polarizations in terms of the components of the Riemann tensor governing the geodesic deviation.

In Sec.~\ref{sec:mg}, we formulated the polarizations in linear MG with generic, non-Fierz--Pauli-type masses.
We identified all the independent variables that obey Klein--Gordon-type equations.
The dynamical dofs in the generic MG consist of spin-2 and spin-$ 0 $ modes;
the former breaks down into two tensor (helicity-2), two vector (helicity-$ 1 $) and one scalar (helicity-$ 0 $) polarizations, while the latter just corresponds to a scalar polarization.
We found convenient ways of decomposing the two scalar modes of each spin into distinct linear combinations of the transverse and longitudinal polarizations as in \eqref{eq:Riemann_mg}.
This expression contains the graviton masses as the coefficients, so we expect it will serve as a useful tool in measuring the masses of GWs.

In Sec.~\ref{sec:hcg}, we analyzed the linear perturbations of generic HCG whose Lagrangian is an arbitrary polynomial of the Riemann tensor.
When expanded around a flat background, the linear dynamical dofs in this theory are identified as massless spin-$ 2 $, massive spin-$ 2 $ and massive spin-$ 0 $ modes.
The massive spin-$ 2 $ arises from the Weyl-squared term in the Lagrangian and the massive spin-$ 0 $ from the Ricci scalar squared.
The massless spin-$ 2 $ is characterized by the tensor-type (helicity-$ 2 $) polarization modes.
As its massive part encompasses the identical structure to the generic MG, GWs in the generic HCG provide six massive polarizations on top of the ordinary two massless modes.
In parallel to MG, we found convenient representations for the scalar polarizations directly connected to the coupling constants of HCG as in \eqref{eq:Riemann_hcg}.

In the analysis of HCG, we used two methods and showed that the two results agree.
One takes full advantage of the partial equivalence between the generic HCG and MG at the linear level, whereas the other relies upon a gauge-invariant formalism originally developed for cosmological perturbation theories.
The present result about the scalar part can be compared with the case of inflationary cosmological perturbations in Einstein--Weyl gravity studied in \cite{Deruelle:2010kf}, where the conformal analogue of the gauge-invariant variable $ W = \Psi-\Phi $ becomes dynamical.

In Sec.~\ref{sec:obs}, we gave a brief discussion about possible methods to determine the theory parameters by means of GW-polarization measurements with emphasis on the merit that they do not require measuring the propagation speeds, whether absolute or relative to other signals, or the details of the waveforms of the GWs.
In any case, the full development in this direction is left to future work.

\begin{acknowledgments}
The authors are grateful to Yuki Niiyama for fruitful discussions.
They thank Hideki Asada for useful comments.
\end{acknowledgments}

\appendix

\section{\label{sec:gauge}Gauge transformations and gauge-invariant variables}

In order to construct gauge-invariant variables, let us consider an active transformation of the coordinate system under which the coordinates of any point change according to
\begin{equation}
x^\mu
\rightarrow
  x^\mu + \xi^\mu(x)\,,
\end{equation}
where the vector field $ \xi^\mu $ is as small as the perturbation.
Accordingly, the space-time metric transforms as
\begin{equation}
g_{\mu\nu}
\rightarrow
  g_{\mu\nu} - \pounds_\xi g_{\mu\nu}
\end{equation}
where the arrow denotes the transformation induced by the coordinate change and $ \pounds_\xi $ the Lie derivative along $ \xi^\mu $\,.
It follows that
\begin{equation}
h_{\mu\nu}
\rightarrow
  h_{\mu\nu} - \pounds_\xi \eta_{\mu\nu}
\end{equation}
to first order in perturbations.
The vector field $ \xi^\mu $ can be decomposed into the scalar and vector parts as
\begin{equation}
(\xi^\mu)
= (T,\partial^i L + L^i)
\end{equation}
with $ \partial_i L^i = 0 $.
It is obvious that this does not affect the tensor variable:
\begin{equation}
H_{ij}
\rightarrow
  H_{ij}\,.
\end{equation}
On the other hand, the vector variables are transformed as
\begin{equation}
B_i
\rightarrow
  B_i + \dot L_i\,,
\quad
E_i
\rightarrow
  E_i - L_i\,,
\end{equation}
so the following combination is found to be invariant:
\begin{equation}
\Sigma_i
\equiv
  B_i + \dot E_i\,.
\end{equation}
The transformations of the scalar variables are
\begin{equation}
A
\rightarrow A
  -\dot T\,,
\quad
B
\rightarrow
  B - T + \dot L\,,
\quad
C
\rightarrow
  C\,,
\quad
E
\rightarrow
  E - L\,,
\end{equation}
from which a useful set of invariant combinations is found to be
\begin{equation}
\Psi
\equiv
  A - \dot B - \ddot E\,,
\quad
\Phi
\equiv
  C\,.
\end{equation}

\section{\label{sec:pert}Expressions for curvature tensors and higher-curvature Lagrangian}

This Appendix summarizes the necessary expressions for the linear-order curvature in the Minkowski background as well as the quadratic-curvature action integrals expanded up to second order.
Thanks to the topological nature of the Gauss--Bonnet combination in four dimensions, the Weyl-squared action can be rewritten as
\begin{equation}
S_C
\equiv
  \frac{-\alpha}{2\kappa}\,
  \int\!\mathrm d^4x\,\sqrt{-g}\,C_{\mu\nu\rho\sigma}\,C^{\mu\nu\rho\sigma}
= \frac{-\alpha}{2\kappa}\,
  \int\!\mathrm d^4x\,\sqrt{-g}\,\left(2 R_{\mu\nu}\,R^{\mu\nu} - \frac{2}{3}\,R^2\right)
\end{equation}
up to irrelevant surface integrals.
Thus, to compute its second-order expansion, we only need the first-order Ricci tensor
\begin{equation}
\begin{aligned}
{}^{(1)}R_{\mu\nu}
&
= -\frac{1}{2}\,\square h_{\mu\nu}
  + \partial^\alpha \partial_{(\mu} h_{\nu)\alpha}
  - \frac{1}{2}\,\partial_\mu \partial_\nu h\,, \\
{}^{(1)}R_{ij}
&
= -\square H_{ij}
  + \partial_{(i} \dot\Sigma_{j)}
  - \partial_i \partial_j \Psi
  - \partial_i \partial_j \Phi
  - \delta_{ij}\,\square \Phi\,, \\
{}^{(1)}R_{i0}
&
= \frac{1}{2}\,\triangle \Sigma_i
  - 2 \partial_i \dot\Phi\,, \\
{}^{(1)}R_{00}
&
= \triangle \Psi
  - 3 \ddot\Phi
\end{aligned}
\end{equation}
and the Ricci scalar
\begin{equation}
{}^{(1)}R
= \partial_\mu \partial_\nu h^{\mu\nu} - \square h
= -2 \triangle(\Psi-\Phi)
  - 6 \square\Phi\,.
\end{equation}
The Weyl-squared action expanded up to second order is given in terms of the perturbative variables as
\begin{equation}
\begin{aligned}
{}^{(2)}S_C
&
= \frac{-\alpha}{2\kappa}\,\int\!\mathrm d^4x\,\left(
   2 {}^{(1)}R_{\mu\nu}\,{}^{(1)}R^{\mu\nu}
   - \frac{2}{3}\,{}^{(1)}R^2
  \right) \\
&
= \frac{-\alpha}{2\kappa}\,\int\!\mathrm d^4x\,
  \left[
   \frac{1}{2}\,\square h_{\mu\nu}\,\square h^{\mu\nu}
   - \square h_{\mu\nu}\,\partial_\alpha \partial^\mu h^{\alpha\nu} 
   + \frac{1}{3}\,\square h\,\partial_\mu \partial_\nu h^{\mu\nu}
   + \frac{1}{3}\,(\partial_\mu \partial_\nu h^{\mu\nu})^2 
   - \frac{1}{6}\,(\square h)^2
  \right] \\
&
= \frac{-\alpha}{2\kappa}\,\int\!\mathrm d^4x\,
  \left[
   2 (\square H_{ij})^2
   + (\partial_i \dot\Sigma_j)^2
   - (\triangle\Sigma_i)^2
   + \frac{4}{3}\,\left[\triangle (\Psi-\Phi)\right]^2
  \right]\,,
\end{aligned}
\end{equation}
where surface terms have been discarded.
The computation of the second-order expansion of the Ricci-squared action
\begin{equation}
S_R
\equiv
  \frac{\beta}{2\kappa}\,\int\!\mathrm d^4x\,\sqrt{-g}\,R^2
\end{equation}
is straightforward:
\begin{equation}
\begin{aligned}
{}^{(2)}S_R
&
= \frac{\beta}{2\kappa}\,\int\!\mathrm d^4x\,{}^{(1)}R^2 \\
&
= \frac{\beta}{2\kappa}\,\int\!\mathrm d^4x\,
  \left[
   (\partial_\mu \partial_\nu h^{\mu\nu})^2 
   - 2 \square h\,\partial_\mu \partial_\nu h^{\mu\nu}
   + (\square h)^2
  \right] \\
&
= \frac{\beta}{2\kappa}\,\int\!\mathrm d^4x\,
  4 \left[\triangle (\Psi-\Phi) + 3 \square\Phi\right]^2\,.
\end{aligned}
\end{equation}

\section{\label{sec:dec}Decoupling dofs in quadratic curvature gravity on arbitrary Einstein manifolds}

In this Appendix, we describe the equivalence of the quadratic curvature gravity (QCG) and GR ``minus'' MG at the linear level on arbitrary Einstein manifolds \cite{Niiyama:2019fvf}.
Let us begin with the generic quadratic curvature action with a cosmological constant
\begin{equation}
S_\mathrm{QCG}[g_{\mu\nu}]
= \frac{1}{2\kappa}\,\int\!\mathrm d^4x\,\sqrt{-g}\,\left(
   R
   - 2 \Lambda
   - \alpha\,C_{\mu\nu\rho\sigma}\,C^{\mu\nu\rho\sigma}
   + \beta\,R^2
  \right)\,,
\end{equation}
where we have dropped the topological Gauss--Bonnet term.
This theory admits any metric $ \bar g_{\mu\nu} $ satisfying
\begin{equation}
R_{\mu\nu}[\bar g_{\mu\nu}]
= \Lambda\,\bar g_{\mu\nu}
\end{equation}
as a solution to the eom.
It is useful to introduce a Lovelock tensor
\begin{equation}
\mathcal G_{\mu\nu}
\equiv
  G_{\mu\nu} + \Lambda\,g_{\mu\nu}\,,
\end{equation}
which vanishes when evaluated with $ \bar g_{\mu\nu} $\,.
The action can be rewritten as
\begin{equation}
S_\mathrm{QCG}[g_{\mu\nu}]
= \frac{\chi}{2\kappa}\,\int\!\mathrm d^4x\,\sqrt{-g}\,\left(
   2 \Lambda
   - \mathcal G
   - \frac{\tilde\alpha}{2}\,\mathcal G_{\mu\nu}\,\mathcal G^{\mu\nu}
   + \frac{\tilde\beta}{2}\,\mathcal G^2
  \right)\,,
\end{equation}
where $ \mathcal G = g^{\mu\nu}\,\mathcal G_{\mu\nu} $\,, $ \chi \equiv 1 + (8\beta+4\alpha/3)\,\Lambda $, $ \tilde\alpha \equiv 4\alpha/\chi $ and $ \tilde\beta \equiv (2\beta+4\alpha/3)/\chi $ and where we have again discarded the Gauss--Bonnet term.
Taking $ \bar g_{\mu\nu} $ as the background and expanding the action up to quadratic order in $ h_{\mu\nu} \equiv g_{\mu\nu} - \bar g_{\mu\nu} $\,, we obtain the second-order action for $ h_{\mu\nu} $
\begin{equation}
{}^{(2)}S_\mathrm{QCG}[h_{\mu\nu}]
= \frac{\chi}{4\kappa}\,\int\!\mathrm d^4x\,\sqrt{-\bar g}\,\left(
   -h^{\mu\nu}\,{}^{(1)}\mathcal G_{\mu\nu}[h_{\mu\nu}]
   - \tilde\alpha\,{}^{(1)}\mathcal G_{\mu\nu}[h_{\mu\nu}]\,{}^{(1)}\mathcal G^{\mu\nu}[h_{\mu\nu}]
   + \tilde\beta\,{}^{(1)}\mathcal G[h_{\mu\nu}]^2
  \right)\,,
\label{eq:action_qcg}
\end{equation}
where
\begin{equation}
{}^{(1)}\mathcal G_{\mu\nu}[h_{\mu\nu}]
\equiv
  {}^{(1)}G_{\mu\nu}[h_{\mu\nu}]
  + \Lambda\,h_{\mu\nu}
\end{equation}
and $ {}^{(1)}\mathcal G[h_{\mu\nu}] \equiv \bar g^{\mu\nu}\,{}^{(1)}\mathcal G_{\mu\nu}[h_{\mu\nu}] $.
As usual, the tensor indices are raised and lowered with the background metric $ \bar g_{\mu\nu} $\,.
Replacing $ {}^{(1)}\mathcal G_{\mu\nu}[h_{\mu\nu}] $ in \eqref{eq:action_qcg} with an auxiliary variable $ A_{\mu\nu} $ and adding a constraint leads to
\begin{equation}
{}^{(2)}S_\mathrm{QCG}[h_{\mu\nu},A_{\mu\nu},\lambda_{\mu\nu}]
= \frac{\chi}{4\kappa}\,\int\!\mathrm d^4x\,\sqrt{-\bar g}\,\left(
   -h^{\mu\nu}\,A_{\mu\nu}
   - \tilde\alpha\,A_{\mu\nu}\,A^{\mu\nu}
   + \tilde\beta\,A^2
   + \lambda^{\mu\nu}\,(A_{\mu\nu} - {}^{(1)}\mathcal G_{\mu\nu}[h_{\mu\nu}])
  \right)\,,
\end{equation}
where $ A \equiv \bar g^{\mu\nu}\,A_{\mu\nu} $ and $ \lambda^{\mu\nu} $ is a Lagrange multiplier.
The variation of the above action with respect to $ A_{\mu\nu} $ gives an algebraic constraint
\begin{equation}
\lambda_{\mu\nu}
= h_{\mu\nu}
  + 2 \tilde\alpha\,A_{\mu\nu}
  - 2 \tilde\beta\,A\,\bar g_{\mu\nu}\,,
\end{equation}
which can be substituted back to the action to eliminate $ A_{\mu\nu} $ to give
\begin{equation}
{}^{(2)}S_\mathrm{QCG}[h_{\mu\nu},\lambda_{\mu\nu}]
= \frac{\chi}{4\kappa}\,\int\!\mathrm d^4x\,\sqrt{-\bar g}\,\left[
   -\lambda^{\mu\nu}\,{}^{(1)}\mathcal G_{\mu\nu}[h_{\mu\nu}]
   + \frac{m^2}{8}\,\left(
      (h_{\mu\nu}-\lambda_{\mu\nu})\,(h^{\mu\nu}-\lambda^{\mu\nu})
      - (1-\epsilon)\,(h-\lambda)^2
     \right)
  \right]\,,
\end{equation}
where $ \epsilon \equiv 1 + \tilde\beta/(\tilde\alpha-4 \tilde\beta) = 9\beta/(2\alpha+12\beta) $, $ m^2 \equiv 2/\tilde\alpha = \chi/(2\alpha) $ and $ \lambda \equiv \bar g^{\mu\nu}\,\lambda_{\mu\nu} $\,.
Finally, by transforming
\begin{equation}
h_{\mu\nu}
\to
  \phi_{\mu\nu} + \tilde\phi_{\mu\nu}\,,
\quad
\lambda_{\mu\nu}
\to
  \phi_{\mu\nu} - \tilde\phi_{\mu\nu}\,,  
\end{equation}
we arrive at
\begin{equation}
{}^{(2)}S_\mathrm{QCG}[\phi_{\mu\nu},\tilde\phi_{\mu\nu}]
= \frac{\chi}{4\kappa}\,\int\!\mathrm d^4x\,\sqrt{-\bar g}\,\left[
   -\phi^{\mu\nu}\,{}^{(1)}\mathcal G_{\mu\nu}[\phi_{\mu\nu}]
   + \tilde\phi^{\mu\nu}\,{}^{(1)}\mathcal G_{\mu\nu}[\tilde\phi_{\mu\nu}]
   + \frac{m^2}{2}\,\left(
      \tilde\phi_{\mu\nu}\,\tilde\phi^{\mu\nu}
      - (1-\epsilon)\,\tilde\phi^2
     \right)
  \right]\,,
\end{equation}
where $ \tilde\phi \equiv \bar g^{\mu\nu}\,\tilde\phi_{\mu\nu} $\,.
Equation~\eqref{eq:action_hcg} is obtained as the Minkowski version of this.

\section{\label{sec:noGR}$ \chi = 0 $}

This class contains conformal gravity and $ R^2 $ gravity.
At first glance one might expect this is the massless limit, but it is not.
The action is
\begin{align}
S_\mathrm{HCG}^{(\mathrm T)}[H_{ij}]
&
= \frac{1}{2\kappa}\,\int\!\mathrm d^4x\,\left[
   -2 \alpha\,\Box H_{ij}\,\Box H^{ij}
  \right]\,, \\
S_\mathrm{HCG}^{(\mathrm V)}[\Sigma_i]
&
= \frac{1}{2\kappa}\,\int\!\mathrm d^4x\,\left[
   -\alpha\,\left(
    \partial_j \dot\Sigma_i\,\partial^j \dot\Sigma^i
    - \triangle \Sigma_i\,\triangle \Sigma^i
   \right)
  \right]\,, \\
S_\mathrm{HCG}^{(\mathrm S)}[\Phi,\Theta,\Xi]
&
= \frac{1}{2\kappa}\,\int\!\mathrm d^4x\,\left[
   -3 \alpha\,\Theta^2
   - \beta\,\Xi^2
   - 12 \beta\,\Xi\,\square\Phi
   + 6 \beta\,\Xi\,\Theta
  \right]\,,
\end{align}
where we have already introduced $ \Xi $ to replace $ {}^{(1)}R $.

The eoms for the tensor and vector variables can be easily found:
\begin{equation}
\alpha\,\square^2 H_{ij}
= 0\,,
\quad
\alpha\,\square \triangle \Sigma_i
= 0\,.
\end{equation}
When $ \alpha \neq 0 $, these eoms admit plane-wave solutions
\begin{equation}
H_{ij}
= A_{ij}\,\mathrm e^{\mathrm i\,\omega_A\,(z-t)}
  + t\,B_{ij}\,\mathrm e^{\mathrm i\,\omega_B\,(z-t)}\,,
\quad
\Sigma_i
= C_i\,\mathrm e^{\mathrm i\,\omega_C\,(z-t)}\,,
\end{equation}
with $ A_{ij} $\,, $ B_{ij} $ and $ C_i $ arbitrary constants.
The tensor wave indicates the emergence of an instability.

The scalar part is more involved.
The variations of the action with respect to each scalar variable give a set of equations
\begin{equation}
\beta\,\square\Xi
= 0\,,
\quad
-6 \beta\,\square\Phi
+ 3 \beta\,\Theta
- \beta\,\Xi
= 0\,,
\quad
\alpha\,\Theta - \beta\,\Xi
= 0\,.
\end{equation}
If $ \beta = 0 $, then we have $ \Theta = 0 $ hence $ \Phi = \Psi $, but these cannot be determined.
If $ \beta \neq 0 $ but $ \alpha = 0 $, then $ \Xi = 0 $ and one cannot determine $ \Phi $ and $ \Theta $.
If both $ \alpha $ and $ \beta $ are nonzero but $ \alpha = 3 \beta $, we obtain equations for $ \Phi $ and $ \Psi $ as
\begin{equation}
\square \Phi
= \square \triangle\Psi
= 0\,.
\end{equation}
Finally, in the most generic case when both $ \alpha $ and $ \beta $ are nonzero and $ \alpha \neq 3 \beta $, we can eliminate $ \Xi $ and $ \Theta $ to have the eom for $ \Phi $ as
\begin{equation}
\square^2 \Phi
= 0\,,
\end{equation}
which suggests that $ \Phi $ is unstable.

\section{\label{sec:F}Detector responses}

In this Appendix, we just summarize the angular pattern functions for the six polarizations defined by
\begin{equation}
F_\alpha(\boldsymbol\Omega)
\equiv
  \boldsymbol D : \boldsymbol e_\alpha(\boldsymbol\Omega)\,,
\end{equation}
where $ \boldsymbol D $ is the so-called detector tensor, $ \boldsymbol e_\alpha $ the polarization tensor, $ \boldsymbol\Omega $ the unit vector pointing the impinging direction of a GW and the symbol $:$ denotes contraction between tensors.

In the main text, we have kept using an inertial coordinate system such that a gravitational wave propagates in the $ z $ direction.
We call it the gravitational-wave frame and denote its orthonormal basis as $ (\boldsymbol m,\boldsymbol n,\boldsymbol\Omega) $, where $ \boldsymbol\Omega $ is the unit vector along the $ z $ direction.
Note that there is a rotation degree of freedom along the $ \boldsymbol\Omega $ axis which will be denoted as the polarization angle $ \psi $.
The polarization tensors for $ \alpha \in \{+, \times, x, y, \mathrm B, \mathrm L, \mathrm T, \mathbb T\} $ can be written using the unit vectors as
\begin{equation}
\begin{aligned}
\boldsymbol e_+ 
&
= \boldsymbol m \otimes \boldsymbol m
  - \boldsymbol n \otimes \boldsymbol n\,, \\
\boldsymbol e_\times 
&
= \boldsymbol m \otimes \boldsymbol n
  + \boldsymbol n \otimes \boldsymbol m\,, \\
\boldsymbol e_x
&
= \boldsymbol m \otimes \boldsymbol\Omega
  + \boldsymbol\Omega \otimes \boldsymbol m\,, \\
\boldsymbol e_y
&
= \boldsymbol n \otimes \boldsymbol\Omega
  + \boldsymbol\Omega \otimes \boldsymbol n\,, \\
\boldsymbol e_\mathrm B
&
= \boldsymbol m \otimes \boldsymbol m
  + \boldsymbol n \otimes \boldsymbol n\,, \\
\boldsymbol e_\mathrm L
&
= \sqrt 2\,\boldsymbol\Omega \otimes \boldsymbol\Omega\,, \\
\boldsymbol e_\mathrm T
&
= \sqrt{\frac{2}{3}}\,\left(
   \boldsymbol m \otimes \boldsymbol m
   + \boldsymbol n \otimes \boldsymbol n
   + \boldsymbol\Omega \otimes \boldsymbol\Omega
  \right)\,, \\
\boldsymbol e_\mathbb T
&
= \frac{1}{\sqrt 3}\,\left(
   \boldsymbol m \otimes \boldsymbol m
   + \boldsymbol n \otimes \boldsymbol n
   - 2 \boldsymbol\Omega \otimes \boldsymbol\Omega
  \right)\,.
\end{aligned}
\end{equation}

To characterize ground-based interferometers or pulsar timing arrays, we introduce an inertial coordinate system specified by an orthonormal basis $ (\boldsymbol u,\boldsymbol v,\boldsymbol w) $ such that $ \boldsymbol w $ is the upward normal to the Earth's surface.
We call it the detector frame and introduce the usual polar angles $ (\theta,\phi) $ in this frame to point the GW propagation direction.
We then rotate the detector frame $ (\boldsymbol u,\boldsymbol v,\boldsymbol w) $ to $ (\boldsymbol u',\boldsymbol v',\boldsymbol w') $ so that $ \boldsymbol w' $ points toward the GW propagation as shown in Fig.~2 of Ref.~\cite{Nishizawa:2009bf}.
Their relationship is
\begin{equation}
\left\{
\begin{aligned}
\boldsymbol u' 
&
= \cos\theta\,\cos\phi\,\boldsymbol u
  + \cos\theta\,\sin\phi\,\boldsymbol v
  - \sin\theta\,\boldsymbol w\,, \\
\boldsymbol v'
&
= -\sin\phi\,\boldsymbol u
  + \cos\phi\,\boldsymbol v\,, \\
\boldsymbol w'
&
= \sin\theta\,\cos\phi\,\boldsymbol u
  + \sin\theta\,\sin\phi\,\boldsymbol v
  + \cos\theta\,\boldsymbol w\,. \\
\end{aligned}
\right.
\end{equation}
Finally, introducing $ \psi $ as the angle from $ \boldsymbol u' $ to $ \boldsymbol m $ in the plane perpendicular to $ \boldsymbol w' $, we arrive at the relationship between the mediation coordinates and the GW coordinates
\begin{equation}
\left\{
\begin{aligned}
\boldsymbol m
&
= \cos\psi\,\boldsymbol u'
  + \sin\psi\,\boldsymbol v'\,, \\
\boldsymbol n
&
= -\sin\psi\,\boldsymbol u'
  + \cos\psi\,\boldsymbol v'\,, \\
\boldsymbol\Omega
&
= \boldsymbol w'\,.
\end{aligned}
\right.
\end{equation}

Let us consider L-shaped interferometers such as LIGO, Virgo and KAGRA.
The detector tensor in this case is \cite{Nishizawa:2009bf}
\begin{equation}
\boldsymbol D
= \frac{1}{2}\,\left(
   \boldsymbol u \otimes \boldsymbol u 
   - \boldsymbol v \otimes \boldsymbol v
  \right)\,.
\end{equation}
The antenna pattern functions $ F_\alpha $ are calculated as
\begin{equation}
\begin{aligned}
F_+ (\theta,\phi,\psi)
&
= \frac{1}{2}\,\left(1 + \cos^2\theta\right)\,\cos 2\phi\,\cos 2\psi
  - \cos\theta\,\sin 2\phi\,\sin 2\psi\,, \\
F_\times (\theta,\phi,\psi)
&
= -\frac{1}{2}\,\left(1 + \cos^2\theta\right)\,\cos 2\phi\,\sin 2\psi
  - \cos\theta\,\sin 2\phi\,\cos 2\psi\,, \\
F_x(\theta,\phi,\psi)
&
= \sin\theta\,\left(\cos\theta\,\cos 2\phi\,\cos\psi - \sin 2\phi\,\sin\psi\right)\,, \\
F_y(\theta,\phi,\psi)
&
= -\sin\theta\,\left(\cos\theta\,\cos 2\phi\,\sin\psi + \sin 2\phi\,\cos\psi\right)\,, \\
F_\mathrm B(\theta,\phi)
&
= -\frac{1}{2}\,\sin^2\theta\,\cos 2\phi\,, \\
F_\mathrm L(\theta,\phi)
&
= \frac{1}{\sqrt 2}\,\sin^2\theta\,\cos 2\phi\,, \\
F_\mathrm T
&
= 0\,, \\
F_\mathbb T(\theta,\phi)
&
= -\frac{\sqrt 3}{2}\,\sin^2\theta\,\cos 2\phi\,.
\end{aligned}
\end{equation}
The scalar functions are degenerate since they have the same dependence on the angles.

If we consider a pulsar frequency shift as a signal, the detector tensor is \cite{Yunes:2013dva,Lee:2010cg,Qin:2020hfy}
\begin{equation}
\boldsymbol D
= \frac{1}{2}\,\frac{1}{1 + v\,\boldsymbol\Omega \cdot \boldsymbol w}\,
  \boldsymbol w \otimes \boldsymbol w\,,
\end{equation}
where $ \boldsymbol w $ points in the direction of the pulsar.
It was pointed out in Ref.~\cite{Lee:2010cg,Qin:2020hfy} that there should be a modification factor $ v $ representing the subluminal velocity of GWs in the denominator.
In the luminal ($ v = 1 $) case, the antenna pattern functions are
\begin{equation}
\begin{aligned}
F_+(\theta,\psi)
&
= \frac{1}{2}\,\frac{\sin^2\theta}{1+\cos\theta}\,\cos 2\psi\,, \\
F_\times(\theta,\psi)
&
= -\frac{1}{2}\,\frac{\sin^2\theta}{1+\cos\theta}\,\sin 2\psi\,, \\
F_x(\theta,\psi)
&
= -\frac{1}{2}\,\frac{\sin 2\theta}{1+\cos\theta}\,\cos\psi\,, \\
F_y(\theta,\psi)
&
= \frac{1}{2}\,\frac{\sin 2\theta}{1+\cos\theta}\,\sin\psi\,, \\
F_\mathrm B(\theta)
&
= \frac{1}{2}\,\frac{\sin^2\theta}{1+\cos\theta}\,, \\
F_\mathrm L(\theta)
&
= \frac{1}{\sqrt 2}\,\frac{\cos^2\theta}{1+\cos\theta}\,, \\
F_\mathrm T(\theta)
&
= \frac{1}{\sqrt 6}\,\frac{1}{1+\cos\theta}\,, \\
F_\mathbb T(\theta)
&
= \frac{1}{2\,\sqrt 3}\,\frac{1-3\cos^2 \theta}{1+\cos\theta}\,.
\end{aligned}
\end{equation}
The scalar functions are not degenerate.

\bibliography{gwspol}

\end{document}